# Plastic Deformations in Mechanically Strained Single-Walled Carbon Nanotubes

Dolores Bozovic*, M. Bockrath

Department of Physics, Harvard University, Cambridge, MA 02138

Jason H. Hafner, Charles M. Lieber, Hongkun Park

Department of Chemistry and Chemical Biology, Harvard University, Cambridge, MA 02138

M. Tinkham

Department of Physics, Harvard University, Cambridge, MA 02138

*present address: Rockefeller University, New York, NY 10021

**Abstract**

AFM manipulation was used to controllably stretch individual metallic single-walled carbon nanotubes (SWNTs). We have found that SWNTs can sustain elongations as great as 30% without breaking. Scanned gate microscopy and transport measurements were used to probe the effects of the mechanical strain on the SWNT electronic properties, which revealed a strain-induced increase in intra-tube electronic scattering above a threshold strain of ~5-10%. These findings are consistent with theoretical calculations predicting the onset of plastic deformation and defect formation in carbon nanotubes.

Due to their extraordinary mechanical and electronic properties and potential applications in molecular electronics, single-walled carbon nanotubes (SWNTs) have been the focus of much attention during the past ten years[1,2]. SWNTs provide experimental realizations of quasi-one-dimensional conductors, which can be either metallic or semiconducting depending on their radii and chiralities. Although early studies focused on defect-free metallic nanotubes[3-5], several recent experiments have explored the effects of intrinsic structural defects on the electronic properties of SWNTs[6-8]. These defects have also been directly revealed in atomic-resolution scanning tunneling microscope imaging experiments[6], studied using electron transport measurements and scanned gate microscopy[9], and shown to act as resonant scattering centers[7]. In addition, Park et al.[10] have demonstrated that new defects can be induced via ion implantation and examined their effects on transport.

The effects of mechanical distortion on nanotube transport have also been the subject of both theoretical and experimental study. Tombler et al. used an AFM tip to produce small-angle reversible bending and elongation in nanotubes, while measuring the nanotube conduction[11,12]. A number of groups have also performed calculations to theoretically study the effects of large distortions and mechanical strain, the formation of new defects in nanotubes[13-19], and their effects on transport properties[20-22]. In particular, the effect of axial elongation on the mechanical and electronic structure of carbon nanotubes has been the subject of extensive theoretical study. Calculations using tight-binding and local density approximation methods have predicted that tubes should elongate elastically up to a critical strain that depends on the radius and chirality of the tube. For larger strains, introduction of topological defects, such as bond-rotations, becomes energetically favorable[13-19]. According to calculations by Zhang et al.[13], critical elongations fall in the range of 6% (armchair) to 12% (zigzag) for nanotubes with diameters of ~1 nm.

Further study of defect formation is motivated by the importance of their role in the electron transport properties of carbon nanotubes. Previous experimental studies of mechanically induced structural defects have been performed on MWNTs[23] and ropes of SWNTs[24]. Here, we present experimental observations of the effects of mechanical strain on the electronic properties of individual SWNTs. We used an atomic force microscope (AFM) tip to locally move nanotubes[11,25,26] along a SiO$_2$ substrate and elongate them in a controlled fashion. Using the technique of scanned gate microscopy (SGM)[9,27], we studied scattering by intra-tube defects both before and after manipulation. By varying the extent of the strain, we detected increases in electron back-scattering induced by localized longitudinal strain when the strain exceeded a characteristic value of ~5-10%, in agreement with the theoretically predicted range of threshold strains.

The substrates used in these experiments were degenerately doped silicon wafers with 1 μm of SiO$_2$ on top. We studied single-walled carbon nanotubes synthesized by two different methods. Nanotubes grown by the laser arc-discharge technique[28] were dispersed in ethylene dichloride by ultrasonic agitation and deposited onto the silicon substrates. In order to select for single nanotubes, we imaged the samples with tapping-mode AFM and selected only tubes ~1 nm in height. Individual SWNTs synthesized by the chemical vapor deposition method[29,30] were grown directly onto the surface of the silicon wafers.

The nanotube samples were imaged in tapping mode with a Digital Instruments AFM, and the nanotubes' positions were determined relative to a pre-defined grid of markers. After spin-casting a suitable resist bi-layer, we used electron-beam lithography to pattern the positions of electrodes on top of the nanotubes. A 50 Å adhesion layer of Cr and a 450 Å layer of Au was thermally evaporated onto the substrates to form the electrodes. The remaining resist was removed by immersion in acetone. An AFM image of one such device is shown in Fig. 1(a). Devices incorporating metallic SWNTs were then selected, based on the behavior of their conductance vs. the voltage $V_g$ applied to the substrate, which acts as a gate electrode[1-5].

Pushing on the nanotube segment between the leads with an AFM tip[25,26,31] induces mechanical strain in the tube. Before AFM manipulation, we first imaged the tubes in tapping mode. We then set the oscillation amplitude of the tip to zero and disengaged the scanner feedback. Lowering the tip further by ~100 nm brings it into contact with the substrate surface nearby the nanotube. In this state, moving the tip laterally pushes the nanotube along with it. Figs. 1 (a) and (b) show tapping-mode images of a nanotube before and after pushing it in a direction perpendicular to its long axis.

Varying the extent of the lateral tip motion produces controllable levels of axial strain in the carbon nanotube. Imaging the nanotube after such deformation (see Fig. 1(b)) reveals that frictional forces between the tube and the substrate are sufficient to maintain the elongation of the SWNTs after the removal of the AFM tip.

Comparing the topographic images taken before and after straining (left panels of Fig. 2 (a) and (b)), we observed that neither the shape nor the position of the nanotubes is visibly altered by manipulation except near where the AFM tip pushed the nanotube. This indicates that the displacement-induced strain is non-uniform. To obtain an approximate estimate of the local average strain within the observably displaced region, we make two simplifying assumptions: that most of the strain is localized to the region that visibly moved under manipulation, and that it is uniformly distributed within this region. We measure the average strain by comparing the lengths of the SWNT segments that moved visibly before and after straining. The location of these perturbed segments was determined by overlaying the topographic images taken before and after straining, and their respective lengths were digitally measured using a standard drawing software package [Deneba Canvas 5.0]. We found that local average elongations as great as 30% could be achieved without breaking the nanotubes mechanically or electrically, allowing us to study strains well into the theoretically predicted plastic regime. Outside of the maximally strained region, we estimate the typical strain to be $\leq 1\%$, based on the positional measurement resolution in our experiment of ~20 nm.

Several checks were performed to verify that the tubes were indeed elongated in the central segment between the metallic electrodes, rather than sliding along the surface axially. We took topographic AFM images of nanotube devices before and after AFM manipulation, directly comparing the positions of the tube ends. In over 30 samples that we strained in this fashion, we have never observed any sliding in of the nanotube ends. Another possible mechanism of strain release would be nanotubes breaking and sliding directly underneath the gold electrodes. As a check, we strained a number of nanotube samples sufficiently that, had the ends been broken, they would have been pulled out entirely from beneath the leads. This was not observed in any of the samples. Furthermore, we found that nanotubes do eventually break under sufficient strain, but we have always found them to break close to the point where they were pushed by the tip, indicating that, for large tip displacements, the strain is concentrated near the point of manipulation.

To determine whether the elongation altered the electronic properties of the SWNTs, the SWNTs were probed both before and after manipulation using scanned gate microscopy[9,27] (SGM) and transport measurements. In SGM, a voltage is applied to a conductive AFM tip (typically in our experiment 5-10 V), which is used as a moveable gate. Bringing the tip to a position directly above a nanotube modulates the local charge density in the tube and therefore shifts its local Fermi level. Moving the tip further away decreases its capacitive coupling to the nanotube and therefore reduces its gating effect. Plotting the conductance of the tube as a function of the tip position for fixed $V_g$ yields a spatially resolved image of the effects of this modulation.

In our previous work[7], we studied scattering of electrons by intrinsic defects in single-walled carbon nanotubes. As-grown metallic tubes were found to exhibit conductances with non-monotonic dependence on $V_g$. SGM images that were taken on these devices showed ring-like features, which were interpreted as signatures of resonant scattering by intrinsic defects. According to theoretical predictions[20], these defects (for example, vacancies) lead to quasi-bound defect states, which in turn give rise to increased scattering at specific energies. Both types of nanotubes used in this experiment, those grown by the laser arc-discharge and those grown by the chemical vapor deposition methods, were found to contain a number of native defects.

Following this measurement on unstrained nanotubes, we used AFM manipulation to induce varying degrees of strain in SWNTs, in the range of ~2-30%, and investigated the resulting impact on the nanotube electronic properties. In devices strained by only a few percent, we did not observe the introduction of new scattering centers in the SGM scans. This indicates that mere contact with the AFM tip is not sufficient to introduce new scattering centers in the nanotube lattice, and hence, the tube elongation is most likely elastic.

For larger strains, we observed qualitatively different behavior. Figure 2 shows topographic and scanned gate images taken on a nanotube device before and after AFM manipulation, where a change in electronic properties occurred. A local elongation of ~7 ± 1% over a length of about 520 nm was induced in this nanotube. The scanned gate image taken before straining showed a number of intrinsic defects. After manipulation, the SGM image corresponding to the maximally strained portion (characterized by having observable displacement) of the nanotube shows one large feature. The SGM image corresponding to the remaining portion of the tube, on the contrary, still shows relatively smaller features corresponding to individual scattering centers. These are qualitatively similar in appearance to ones shown in Fig. 2(a), although the individual scatterers show increased resistances as compared to the unstrained tube. Thus we find that the perturbation caused by the tip affects transport properties of the SWNT differently, depending on whether it was scanned over the maximally strained or the remaining segments respectively. This also indicates that the AFM manipulation modulates the electronic properties of the SWNT along the entire observably displaced region.

The resistance of this sample as a function of $V_g$ is shown in Fig. 3. The plots show $V_g$ sweeps taken before and after deformation of the nanotube. Both scans show gate modulation characteristic of tubes with defects[7], consistent with the appearance of rings in the corresponding before and after SG images, with a peak in the resistance occurring at positive $V_g$. As can be seen from these two scans, the resistance of the nanotube for $V_g$ where the defects scatter most strongly (maximum in the resistance rise) has increased by a factor of ~20 after the induced elongation. On the other hand, when $V_g$ is tuned away from the maximum resistance value (negative $V_g$) the resistance had changed only by a small amount.

The large increase in the nanotube resistance occurring selectively at particular values of $V_g$, together with the local change in electronic properties of the strained SWNT segment evinced by the SGM images, shows that the maximally strained SWNT portion shows an increase in back-scattering that is resonant in nature. Overall, for samples strained above 10%, comparing the resistances and SGM images taken before and after straining revealed similar behavior to the sample shown above. In the intermediate regime, from ~5-10% strain, some samples showed an increase in local back-scattering, while others did not. Samples elongated less than 5% showed little change in behavior.

Theoretical calculations predict that the critical strains for defect formation depend on nanotube chiralities[13], falling in the range from 6 to 12%. Although we do not know the detailed atomic structure of our nanotubes, we have induced a wide range of elongations, and observed dramatic increases in local back-scattering for strains higher than ~5-10%, which is consistent with these theoretical predictions for the onset of plastic deformation and defect formation. For each of our samples, we also measured the nanotube diameters from the topographic AFM scans. We observed no notable dependence of the electronic scattering properties on diameter in the measured range of

~1-3 nm. Finally, we note that the changes evident in the SGM images corresponding to the portions of the SWNT outside the maximally strained region may result from the distortion of intrinsic defects, causing them to scatter more strongly. Further theoretical and experimental work is required to fully characterize and understand this effect.

In conclusion, we have used an atomic force microscope tip to induce varying degrees of strain in single-walled carbon nanotubes. Due to the robustness of the nanotube lattice, strains as great as 30% can be reproducibly induced, bringing the nanotube into the theoretically anticipated regime of plastic deformation. Transport measurements and scanned gate microscopy were then used to study scattering in nanotubes before and after manipulation and to thereby observe the induced increase in local resonant back-scattering. Elucidation of the threshold of defect formation in nanotubes, besides providing insight into their structure, is an important step towards their potential applications as strengthening materials and as miniature electro-mechanical devices.

This research was supported in part by NSF grants DMR-0072618 and PHY-0117795.

[31]J. Lefebvre, J.F. Lynch, M. Llaguno, M. Radosavljevic, and A.T. Johnson, Appl. Phys. Lett. **75,** 3014 (1999).

**Figure captions**

FIG. 1. Differential tapping-mode AFM color-enhanced images were taken (a) before and (b) after straining the nanotube with the AFM tip. In this device, Cr/Au leads had been deposited on top of the SWNT (the underlying tube formed small ridges in the surface of the metallic leads). As the nanotubes were anchored to the surface by the gold electrodes, this manipulation forced the central portion of the tube to elongate.

FIG 2. (a) The top panel gives topographic (left) and SGM (right) images taken on a nanotube device before AFM manipulation. The data scale in this SGM image has been adjusted to correspond to a range from 20-30 μS, darkest to lightest areas respectively. (b) The bottom panel shows images taken on the same device after straining. The data scale in this image ranges from 0.2 to 15 μS. We used two different conductive tips to acquire these two images, with a tip voltage of 10V applied to both.

FIG 3. Resistance is plotted as a function of the $V_g$, measured on the device shown in Figure 2. The curves show scans measured (a) before and (b) after straining the nanotube. Please note the difference in scales between the two plots. The peaks in the resistance correspond to maximal scattering by the induced scattering centers, as tuned by varying $V_g$. The graphs indicate that the nanotube resistance had not changed significantly in absolute amount in the region of negative $V_g$, whereas it had increased by a factor of ~20 in the regimes where the centers have been tuned to scatter maximally. Both measurements were taken at room temperature.

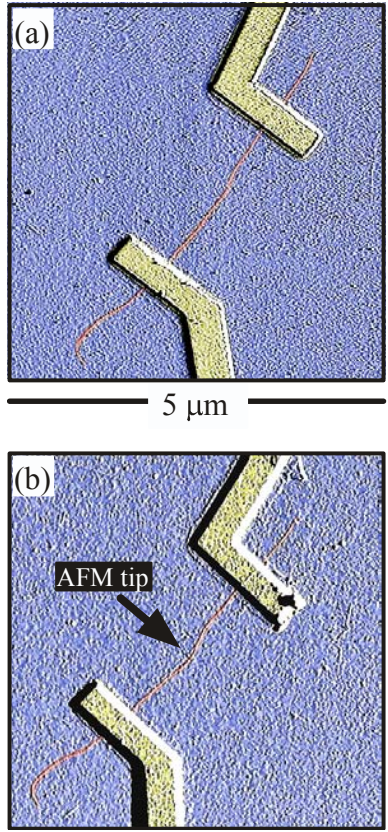

Bozovic et al., Figure 2

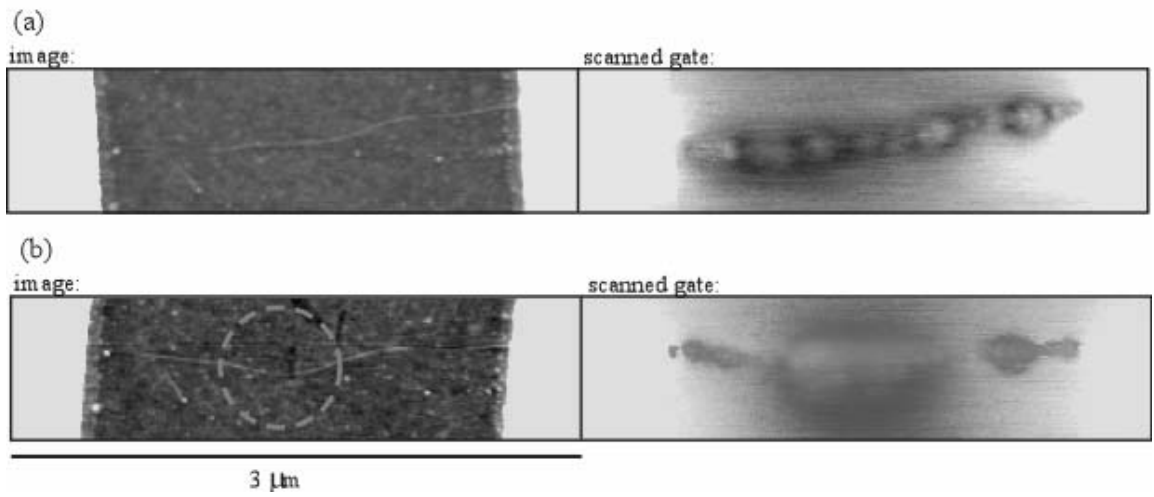

Bozovic et al., Figure 3

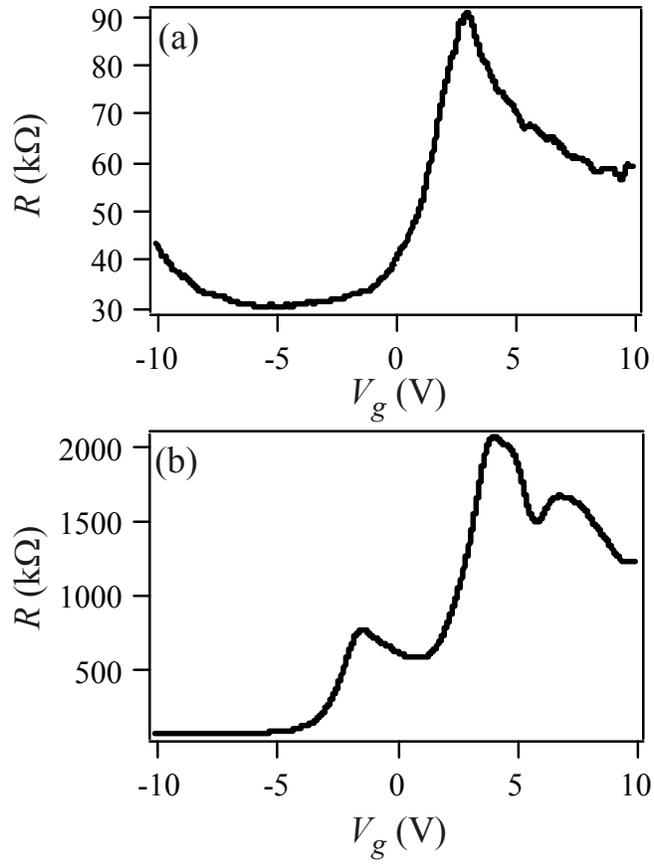